\documentclass[12pt]{iopart}
\usepackage[dvipdfmx]{graphicx}
\usepackage{bm}
\usepackage{comment}
\usepackage{color}

\begin{document}

\title[]
{Non-unique detailed constructions of Curzon-Ahlborn cycle on thermodynamic plane}

\author{Yuki Izumida}

\address{Department of Complexity Science and Engineering, Graduate School of Frontier Sciences, The University of Tokyo, Kashiwa 277-8561, Japan}
\ead{izumida@k.u-tokyo.ac.jp}
\vspace{10pt}
%\begin{indented}
%\item[]September 2023
%\end{indented}

\begin{abstract}
The Curzon-Ahlborn (CA) cycle is a paradigmatic model of endoreversible heat engines, which yields the so-called CA efficiency as the efficiency at maximum power.
Due to the arbitrariness of the relationship between the steady temperature and the time taken for the isothermal process of the CA cycle, the constructions of the CA cycle on the thermodynamic plane are not unique. Here, we give some of the detailed constructions of the CA cycle on the thermodynamic plane, 
using an ideal gas as a working substance.
It is shown that these constructions are equal to each other in the maximum power regime
in the sense that they achieve the best trade-off between the work and the inverse cycle-time, known as the Pareto front in multi-objective optimization problems.
\end{abstract}

%\maketitle

\section{Introduction}
Since the discovery by Carnot, it is known that the efficiency $\eta$ of any heat engine operating with hot and cold heat baths is bound from above by the Carnot efficiency:
\begin{eqnarray}
\eta \le 1-\frac{T_c}{T_h},
\end{eqnarray}
where $T_h$ and $T_c$ denote the temperatures of the hot and cold heat baths, respectively.
The Carnot cycle is an idealized reversible cycle that attains the Carnot efficiency, which is constituted with isothermal processes and adiabatic processes connecting the isothermal processes.
The Carnot cycle can be expressed as a pressure-volume diagram on the thermodynamic plane, which was introduced by Clapeyron.
The analysis of the Carnot cycle and thermodynamics itself was much advanced by the introduction of such a diagram.

As the Carnot cycle is operated quasistatically to meet the reversible condition, its power, the work output divided by the cycle time, vanishes.
Heat engines should run a cycle in a finite time to output a finite power.
So, the power, besides the efficiency, also characterizes the performance of heat engines.
In 1975, Curzon and Ahlborn proposed a simple model of a finite-time Carnot cycle (CA cycle) and studied its efficiency~\cite{CA1975}.
Specifically, they derived the following formula as the efficiency at maximum power $\eta^*$:
\begin{eqnarray}
\eta^*=1-\sqrt{\frac{T_c}{T_h}}.\label{eq.CA}
\end{eqnarray}
This efficiency was derived only by assuming the Newton' law of cooling as a heat transfer law between the steady temperature of a working substance and that of the heat bath and an endoreversibie condition, where the latter is the condition for the cycle to close. A remarkable feature of eq.~(\ref{eq.CA}) is that it depends only on the ratio of the temperatures of the heat baths and is independent of the other detailed parameters of the model.
It should be noted that eq.~(\ref{eq.CA}) was re-derivation; several authors had already derived essentially the same formula for some of the heat engine models more previously~\cite{Y1955,C1957,N1958} (see also~\cite{B1996,VLF2014,MP2015} for historical backgrounds).
Considering the generality of the model and its impact on the subsequent studies, for brevity, we call eq.~(\ref{eq.CA}) the CA efficiency in the present paper.

In fact, the paper by Curzon and Ahlborn opened up a new avenue of discipline called finite-time thermodynamics~\cite{ASB1984,ABOS1984}.
Furthermore, for almost two decades, the determination of the efficiency at maximum power of heat engines, specifically focusing on the universality of the CA efficiency,  has been one of the central issues in nonequilibrium statistical mechanics and thermodynamics, which includes the linear response regimes~\cite{VdB2005,JCH2007} and beyond~\cite{ELVdB2009,EKLVdB2010}.

With these backgrounds, here are some of the key questions that constitute the main subject of this paper:
What does it look like when we illustrate the CA cycle on the thermodynamic plane as we do for the Carnot cycle? Is it unique? 
Where should we switch the volume of each thermodynamic process for the next thermodynamic process in the CA cycle? These questions are especially important when designing the CA cycle using a specific working substance, whose answers are not found in the paper by Curzon and Ahlborn~\cite{CA1975}.
Although some explicit constructions that mimic the CA cycle on the thermodynamic plane were proposed~\cite{GKPR1978,LL2002,IO2015,IO2017,RGRGAAB2018}, 
the above questions have not yet been answered satisfactorily.

The purpose of this paper is to answer the above questions by giving detailed constructions of the CA cycle by expressing it on the thermodynamic plane using an ideal gas as a working substance.
Due to the arbitrariness of the relationship between the steady temperature and the duration taken for the isothermal process of the CA cycle, the constructions of the CA cycle on the thermodynamic plane are not unique. Therefore, we give the simplest construction of the CA cycle first and compare it with other, more complicated constructions.
These constructions, however, are shown to be equal to each other in the maximum power regime by employing the concept of the Pareto front in multi-objective optimization problems~\cite{D2001} in which the best trade-off between the work and the inverse cycle-time regarded as the speed of the cycle is achieved.

\section{Curzon-Ahlborn cycle}
Before presenting the main results in the next section, we review the derivation of the CA efficiency based on~\cite{CA1975} and discuss its properties.
Let $T_{hw}$ and $T_{cw}$ denote the steady temperatures of the working substance during the isothermal and compression processes of the CA cycle, which last for the times $t_h$ and $t_c$, respectively.
In the CA cycle, the heat transfers per unit time $J_h$ and $J_c$ are assumed to obey the Newton's law of cooling:
\begin{eqnarray}
&J_h=\alpha (T_h-T_{hw}),\label{eq.qh}\\
&J_c=\beta (T_{cw}-T_c),\label{eq.qc}
\end{eqnarray}
where $\alpha$ and $\beta$ denote the heat-transfer conductances of the isothermal expansion and processes, respectively.
In addition, the following endoreversible condition~\cite{CA1975,R1979} as the condition for the cycle to close is assumed:
\begin{eqnarray}
\frac{Q_h}{T_{hw}}-\frac{Q_c}{T_{cw}}=0,\label{eq.endo}
\end{eqnarray}
where $Q_h\equiv J_ht_h$ and $Q_c\equiv J_ct_c$ denote the total heats transferred during the isothermal expansion and compression processes, respectively.
By defining the work per cycle as $W\equiv Q_h-Q_c$, the efficiency $\eta$ of the CA cycle reads
\begin{eqnarray}
\eta=\frac{W}{Q_h}=1-\frac{Q_c}{Q_h}=1-\frac{T_{cw}}{T_{hw}},\label{eq.effi_def}
\end{eqnarray}
where we used the endoreversible condition eq.~(\ref{eq.endo}) in the second equality.

The power of the cycle $P$ is given by
\begin{eqnarray}
P&=\frac{\alpha(T_h-T_{hw})t_h-\beta(T_{cw}-T_c)t_c}{t_h+t_c}\nonumber\\
&=\frac{\alpha(T_h-T_{hw})(t_h/t_c)-\beta(T_{cw}-T_c)}{(t_h/t_c)+1}.\label{eq.pow_def}
\end{eqnarray}
Here, for simplicity and without loss of generality, we have neglected the time taken for the adiabatic processes compared to the total cycle time~\cite{C1985,CY1989}.
Even so, we still assume that the adiabatic processes may be considered sufficiently close to the quasistatic adiabatic processes.
This assumption can be justified because the equilibration time of the working substance in the adiabatic process is generally much shorter than the time until the equilibrium between the working substance and the heat bath is reached in the isothermal process~\cite{C1985}. 
This amounts to choosing the parameter ``$\gamma$" in the paper by Curzon and Ahlborn~\cite{CA1975} to be unity, in which the time taken for the adiabatic processes is assumed to be proportional to that for the isothermal processes.

From the endoreversible condition eq.~(\ref{eq.endo}), the ratio between the times of the isothermal processes is expressed in terms of $T_{hw}$ and $T_{cw}$ as
\begin{eqnarray}
\frac{t_h}{t_c}=\frac{\beta T_{hw}(T_{cw}-T_c)}{\alpha T_{cw}(T_h-T_{hw})}.\label{eq.time_ratio}
\end{eqnarray}
By putting eq.~(\ref{eq.time_ratio}) into eq.~(\ref{eq.pow_def}), we can maximize the power in terms of $T_{hw}$ and $T_{cw}$.
As a consequence, we obtain the following steady temperatures at the maximum power:
\begin{eqnarray}
&T_{hw}^*=\frac{(\alpha T_h)^{1/2}+(\beta T_c)^{1/2}}{\alpha^{1/2}+\beta^{1/2}}T_h^{1/2},\label{eq.Tst_h}\\
&T_{cw}^*=\frac{(\alpha T_h)^{1/2}+(\beta T_c)^{1/2}}{\alpha^{1/2}+\beta^{1/2}}T_c^{1/2}.\label{eq.Tst_c}
\end{eqnarray}
By putting eqs.~(\ref{eq.Tst_h}) and (\ref{eq.Tst_c}) into eq.~(\ref{eq.effi_def}), we recover the CA efficiency in eq.~(\ref{eq.CA}).
From eqs.~(\ref{eq.pow_def})--(\ref{eq.Tst_c}), the maximum power also reads
\begin{eqnarray}
P^*=\alpha \beta \left [\frac{T_h^{1/2}-T_c^{1/2}}{\alpha^{1/2}+\beta^{1/2}}\right ]^2.\label{eq.Pmax}
\end{eqnarray}
At the maximum power, the time ratio eq.~(\ref{eq.time_ratio}) becomes
\begin{eqnarray}
\frac{t_h^*}{t_c^*}=\frac{\beta^{1/2}}{\alpha^{1/2}}.\label{eq.time_ratio_Pmax}
\end{eqnarray}
In the above derivation, there remains arbitrariness in the CA cycle
in the sense that while the time ratio at the maximum power is determined as eq.~(\ref{eq.time_ratio_Pmax}), the total cycle time itself is undetermined. 
Moreover, the work at the maximum power is also undetermined.
This originates from the fact that the power of the CA cycle is expressed in terms of the time ratio as eq.~(\ref{eq.pow_def}), which depends on the steady temperatures as eq.~(\ref{eq.time_ratio}).
This arbitrariness implies that we may have several detailed constructions of the CA cycle on the thermodynamic plane, which will be demonstrated in the next section.

\section{Main results}
\subsection{The simplest construction of CA cycle}\label{Simplest construction of CA cycle}
\begin{figure}[!t]
\begin{center}
\includegraphics[scale=0.9]{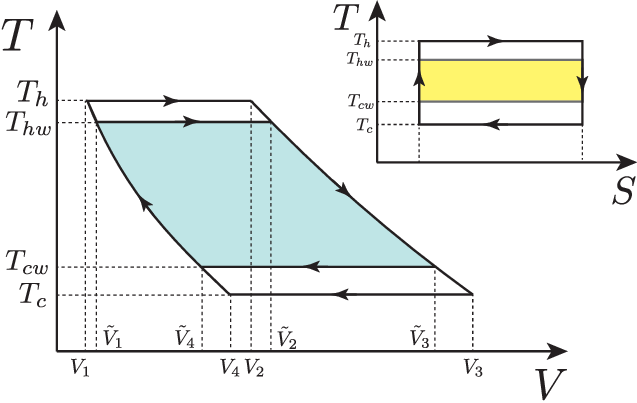}
\caption{The schematic illustration of the temperature-volume ($T$-$V$) diagram of the Carnot cycle (the outer cycle) and the simplest CA cycle $1$ (the shaded inner cycle), where the adiabatic processes of both cycles are made to be overlapped. Upper right: the corresponding temperature-entropy ($T$-$S$) diagram.
The entropy changes during the isothermal processes of the cycle $1$ coincide with the ones of the Carnot cycle.}\label{fig_CA1}
\end{center}
\end{figure}
We give the simplest construction that realizes the CA cycle on the thermodynamic plane by using an ideal gas as the working substance, which we refer to the CA ``cycle $1$" (fig.~\ref{fig_CA1}).

Let $V_i$'s ($i=1, \cdots, 4$) be the switching volumes of the Carnot cycle, which are predetermined in addition to the temperatures $T_h$ and $T_c$ of the heat baths.
Not all $V_i$'s are independent as $V_3$ and $V_4$ are connected by $V_2$ and $V_1$, respectively, through the adiabatic curves as
\begin{eqnarray}
&&T_h V_1^{\gamma-1}=T_c V_4^{\gamma-1},\label{eq.adiabatic_hot}\\
&&T_h V_2^{\gamma-1}=T_c V_3^{\gamma-1},\label{eq.adiabatic_cold}
\end{eqnarray}
where $\gamma$ is the heat capacity ratio.
So, once we determine $V_1$, $V_2$, $T_h$, and $T_c$, $V_3$ and $V_4$ are also determined so that they satisfy eqs.~(\ref{eq.adiabatic_hot}) and (\ref{eq.adiabatic_cold}).

Figure~\ref{fig_CA1} illustrates the temperature-volume ($T$-$V$) diagram of the Carnot cycle and the CA cycle $1$, where the adiabatic processes of these cycles are made to be overlapped~\cite{IO2017,C1985}. Therefore, the switching volumes $V_i$'s of the Carnot cycle and the switching volumes $\tilde V_i$'s of the CA cycle $1$ should satisfy the following relations:
\begin{eqnarray}
&T_hV_1^{\gamma-1}=T_{hw}\tilde V_1^{\gamma-1},\label{eq.tilde_V1}\\
&T_hV_2^{\gamma-1}=T_{hw}\tilde V_2^{\gamma-1},\label{eq.tilde_V2}\\
&T_cV_3^{\gamma-1}=T_{cw}\tilde V_3^{\gamma-1},\label{eq.tilde_V3}\\
&T_cV_4^{\gamma-1}=T_{cw}\tilde V_4^{\gamma-1}.\label{eq.tilde_V4}
\end{eqnarray}
From eqs.~(\ref{eq.tilde_V1})--(\ref{eq.tilde_V4}), we can derive the following relations between $V_i$'s and $\tilde V_i$'s:
\begin{eqnarray}
&\frac{\tilde V_2}{\tilde V_1}=\frac{V_2}{V_1},\label{eq.V1V2_ratio}\\
&\frac{\tilde V_3}{\tilde V_4}=\frac{V_3}{V_4}.\label{eq.V3V4_ratio}
\end{eqnarray}

Denoting the entropy of the working substance during the isothermal expansion and compression processes by $S_h$ and $S_c$, respectively, we may express the heat transfers per unit time eqs.~(\ref{eq.qh}) and (\ref{eq.qc}) in terms of their time-derivatives as follows:
\begin{eqnarray}
&J_h=\alpha(T_h-T_{hw})=T_{hw}\frac{dS_h}{dt},\label{eq.dQ_h}\\
&J_c=\beta(T_{cw}-T_c)=-T_{cw}\frac{dS_c}{dt},\label{eq.dQ_c}
\end{eqnarray}
where
\begin{eqnarray}
\frac{dS_l}{dt}=\left(\frac{\partial S_l}{\partial V}\right)_{T=T_{lw}}\frac{dV}{dt}=\frac{Nk_{\rm B}T_{lw}}{V}\frac{dV}{dt}={\rm const.} \quad (l=h, c)\label{eq.dS}
\end{eqnarray}
for the ideal gas, with $N$ and $k_{\rm B}$ being the particle number and Boltzmann constant, respectively.
By integrating both sides of eqs.~(\ref{eq.dQ_h}) and (\ref{eq.dQ_c}) in time, we have
\begin{eqnarray}
&Q_h=\alpha(T_h-T_{hw})t_h=T_{hw}\Delta S_h=Nk_{\rm B}T_{hw}\ln \frac{\tilde V_2}{\tilde V_1},\label{eq.Qh_1}\\
&Q_c=\beta(T_{cw}-T_c)t_c=-T_{cw}\Delta S_c=Nk_{\rm B}T_{cw}\ln \frac{\tilde V_3}{\tilde V_4},\label{eq.Qc_1}
\end{eqnarray}
where $\Delta S_h$ and $\Delta S_c$ are the entropy changes of the ideal gas during the isothermal expansion and compression processes of the CA cycle $1$, respectively.
Here, it should be noted that $\tilde V_2/\tilde V_1=\tilde V_3/\tilde V_4$ holds from $\Delta S_h+\Delta S_c=0$ as the condition that the CA cycle $1$ closes.
Therefore, the endoreversible condition eq.~(\ref{eq.endo}) is automatically satisfied~\cite{R1979}. 
Because of eqs.~(\ref{eq.V1V2_ratio}) and (\ref{eq.V3V4_ratio}), the entropy changes $\Delta S_h$ and $\Delta S_c$ remain the same as the quasistatic ones
(see the temperature-entropy ($T$-$S$) diagram of the cycle $1$ in fig.~\ref{fig_CA1}).

Using eqs.~(\ref{eq.Qh_1}) and (\ref{eq.Qc_1}), the work output $W\equiv Q_h-Q_c$ is given by
\begin{eqnarray}
W=Nk_{\rm B}(T_{hw}-T_{cw})\ln \frac{\tilde V_2}{\tilde V_1}=Nk_{\rm B}(T_{hw}-T_{cw})\ln \frac{V_2}{V_1},\label{eq.W_1}
\end{eqnarray}
where we used eq.~(\ref{eq.V1V2_ratio}) in the second equality.
Note that eq.~(\ref{eq.W_1}) recovers the quasistatic work $W_{\rm qs}\equiv Nk_{\rm B}(T_h-T_c)\ln (V_2/V_1)$ in the quasistatic limit $t_h \to \infty$ and $t_c \to \infty$.

By using eq.~(\ref{eq.dS}), we can rewrite eqs.~(\ref{eq.dQ_h}) and (\ref{eq.dQ_c}) as
\begin{eqnarray}
&T_h-T_{hw}=\frac{Nk_{\rm B}T_{hw}}{\alpha V}\frac{dV}{dt} \quad (0 \le t \le t_h),\label{eq.V_dyna_h}\\
&T_{cw}-T_c=-\frac{Nk_{\rm B}T_{cw}}{\beta V}\frac{dV}{dt} \quad (0\le t \le t_c),\label{eq.V_dyna_c}
\end{eqnarray}
where we set the initial time to $t=0$ for each of the isothermal processes without loss of generality.
By solving eqs.~(\ref{eq.V_dyna_h}) and (\ref{eq.V_dyna_c}) as the differential equations of $V(t)$ with given $T_{hw}$ and $T_{cw}$, we obtain
the time-dependent volume $V(t)$ during the isothermal processes explicitly as
\begin{eqnarray}
&V(t)=A_h\exp \left(\frac{\alpha(T_h-T_{hw})}{Nk_{\rm B}T_{hw}}t\right) \quad (0 \le t \le t_h),\label{eq.V_Tst_h}\\
&V(t)=A_c\exp \left(-\frac{\beta(T_{cw}-T_c)}{Nk_{\rm B}T_{cw}}t\right) \quad (0 \le t \le t_c),\label{eq.V_Tst_c}
\end{eqnarray}
where $A_h$ and $A_c$ are the integral constants to be determined: The integral constant $A_h$ in eq.~(\ref{eq.V_Tst_h}) is given by using eq.~(\ref{eq.tilde_V1}) as
\begin{eqnarray}
A_h=V(0)=\tilde V_1=\left(\frac{T_h}{T_{hw}}\right)^{\frac{1}{\gamma-1}}V_1.
\end{eqnarray}
In a similar manner, $A_c$ in eq.~(\ref{eq.V_Tst_c}) is also given by using eq.~(\ref{eq.tilde_V3}) as
\begin{eqnarray}
A_c=V(0)=\tilde V_3=\left(\frac{T_{c}}{T_{cw}}\right)^{\frac{1}{\gamma-1}}V_3.
\end{eqnarray}

Subsequently, by integrating both sides of eqs.~(\ref{eq.V_dyna_h}) and (\ref{eq.V_dyna_c}) in time, we have
\begin{eqnarray}
&t_h=\frac{Nk_{\rm B}T_{hw}}{\alpha(T_h-T_{hw})}\ln \frac{\tilde V_2}{\tilde V_1}=\frac{Nk_{\rm B}T_{hw}}{\alpha(T_h-T_{hw})}\ln \frac{V_2}{V_1},\label{eq.t_h}\\
&t_c=\frac{Nk_{\rm B}T_{cw}}{\beta(T_{cw}-T_c)}\ln \frac{\tilde V_3}{\tilde V_4}=\frac{Nk_{\rm B}T_{cw}}{\beta(T_{cw}-T_c)}\ln \frac{V_3}{V_4},\label{eq.t_c}
\end{eqnarray}
where we used eqs.~(\ref{eq.V1V2_ratio}) and (\ref{eq.V3V4_ratio}).
It can be easily checked that eqs~(\ref{eq.t_h}) and (\ref{eq.t_c}) satisfy the time ratio in eq.~(\ref{eq.time_ratio}) by noting $V_2/V_1=V_3/V_4$. 
By solving eqs.~(\ref{eq.t_h}) and (\ref{eq.t_c}) with respect to $T_{hw}$ and $T_{cw}$, respectively, we have
\begin{eqnarray}
&T_{hw}=\frac{T_h}{1+\frac{Nk_{\rm B}}{\alpha t_h}\ln \frac{V_2}{V_1}}\label{eq.Thw_t_h},\\
&T_{cw}=\frac{T_c}{1-\frac{Nk_{\rm B}}{\beta t_c}\ln \frac{V_3}{V_4}}.\label{eq.Tcw_t_c}
\end{eqnarray}
By putting eq.~(\ref{eq.Thw_t_h}) into eq.~(\ref{eq.V_Tst_h}) and putting eq.~(\ref{eq.Tcw_t_c}) into eq.~(\ref{eq.V_Tst_c}), 
we can express the time-dependent volume $V(t)$ in eqs.~(\ref{eq.V_Tst_h}) and (\ref{eq.V_Tst_c}) in terms of $t_h$ and $t_c$ as
\begin{eqnarray}
&V(t)=V_1\left(1+\frac{Nk_{\rm B}}{\alpha t_h}\ln \frac{V_2}{V_1}\right)^{\frac{1}{\gamma-1}}\exp \left(\frac{t}{t_h}\ln \frac{V_2}{V_1}\right) \quad (0 \le t \le t_h),\\
&V(t)=V_3\left(1-\frac{Nk_{\rm B}}{\beta t_c}\ln \frac{V_3}{V_4}\right)^{\frac{1}{\gamma-1}}\exp \left(-\frac{t}{t_c}\ln \frac{V_3}{V_4}\right) \quad (0 \le t \le t_c),
\end{eqnarray}
instead of $T_{hw}$ and $T_{cw}$.

As we find from eqs.~(\ref{eq.t_h}) and (\ref{eq.t_c}) or from eqs.~(\ref{eq.Thw_t_h}) and (\ref{eq.Tcw_t_c}), we obtained the explicit and detailed relations between the steady temperatures and the times taken for the isothermal processes,
which are not available in~\cite{CA1975}. At the maximum power, the times for the isothermal processes in eqs.~(\ref{eq.t_h}) and (\ref{eq.t_c}) read
\begin{eqnarray}
&t_h^*=\frac{Nk_{\rm B}T_{hw}^*}{\alpha(T_h-T_{hw}^*)}\ln \frac{V_2}{V_1},\label{eq.t_h_Pmax}\\
&t_c^*=\frac{Nk_{\rm B}T_{cw}^*}{\beta(T_{cw}^*-T_c)}\ln \frac{V_3}{V_4},\label{eq.t_c_Pmax}
\end{eqnarray}
where $T_{hw}^*$ and $T_{cw}^*$ are given in eqs.~(\ref{eq.Tst_h}) and (\ref{eq.Tst_c}).
Moreover, from eq.~(\ref{eq.W_1}), the work at the maximum power $W^*$ reads
\begin{eqnarray}
W^*&=Nk_{\rm B}(T_{hw}^*-T_{cw}^*)\ln \frac{V_2}{V_1}\nonumber\\
&=\frac{(\alpha T_h)^{1/2}+(\beta T_c)^{1/2}}{\alpha^{1/2}+\beta^{1/2}}Nk_{\rm B}(T_h^{1/2}-T_c^{1/2})\ln \frac{V_2}{V_1},\label{eq.W_simplest}
\end{eqnarray}
where we used eqs.~(\ref{eq.Tst_h}) and (\ref{eq.Tst_c}).
The heat during the isothermal process at the maximum power $Q_h^*$ is also given as
\begin{eqnarray}
Q_h^*=T_{hw}^*\Delta S_h=\frac{(\alpha T_h)^{1/2}+(\beta T_c)^{1/2}}{\alpha^{1/2}+\beta^{1/2}}T_h^{1/2}Nk_{\rm B}\ln \frac{V_2}{V_1},\label{eq.Qh_simplest}
\end{eqnarray}
where we used eqs.~(\ref{eq.Tst_h}), (\ref{eq.Tst_c}), (\ref{eq.V1V2_ratio}), and (\ref{eq.Qh_1}).
We recover the CA efficiency from eqs.~(\ref{eq.W_simplest}) and (\ref{eq.Qh_simplest}).

\subsection{Comparison with other constructions}\label{comparison}
\begin{figure}[!t]
\begin{center}
\includegraphics[scale=0.9]{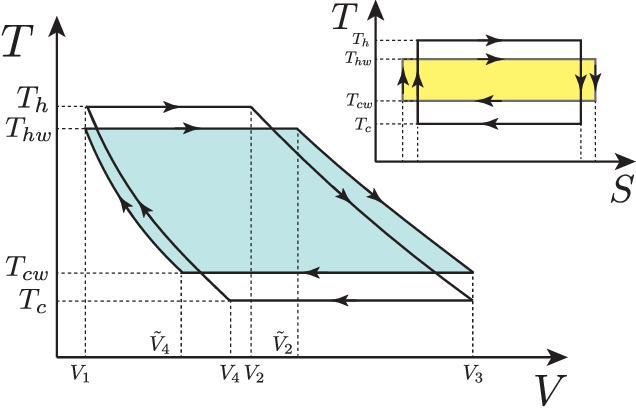}
\caption{The schematic illustration of the temperature-volume ($T$-$V$) diagram of the Carnot cycle (the thinner cycle) and the CA cycle $2$ (the shaded cycle), 
where the minimum volume $V_1$ and the maximum volume $V_3$ are shared with both cycles so that the compression ratio is held fixed. 
Upper right: the corresponding temperature-entropy ($T$-$S$) diagram.
The entropy changes during the isothermal processes of the cycle $2$ are larger than the ones of the Carnot cycle.}\label{fig_CA2}
\end{center}
\end{figure}

\begin{figure}[!t]
\begin{center}
\includegraphics[scale=0.9]{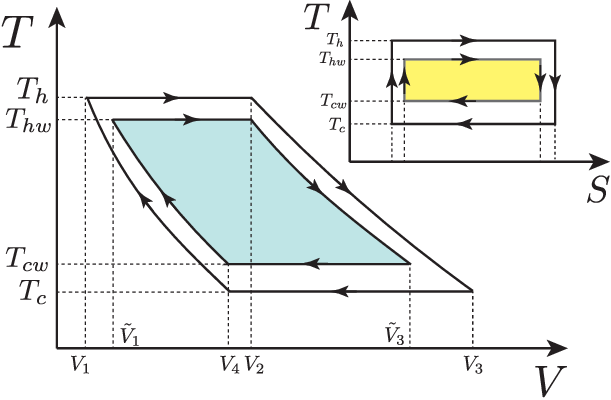}
\caption{The schematic illustration of the temperature-volume ($T$-$V$) diagram of the Carnot cycle (the outer cycle) and the CA cycle $3$ (the inner shaded cycle), 
where $V_2$ and $V_4$ as the end volumes of the isothermal expansion and compression processes, respectively, are shared with both cycles.
Upper right: the corresponding temperature-entropy ($T$-$S$) diagram.
The entropy changes during the isothermal processes of the cycle $3$ are smaller than the ones of the Carnot cycle.}\label{fig_CA3}
\end{center}
\end{figure}

The cycle construction (cycle $1$) introduced in section~\ref{Simplest construction of CA cycle} is not the only construction that yields the CA efficiency as the efficiency at maximum power.
In figs.~\ref{fig_CA2} and \ref{fig_CA3}, we show two cycle constructions that yield the CA efficiency as the efficiency at maximum power, which we refer to the ``cycle $2$" and ``cycle $3$", respectively.
Note that all the CA cycles $1$--$3$ in figs.~\ref{fig_CA1}--\ref{fig_CA3} recover the same Carnot cycle with the predetermined volumes $V_i$'s 
that satisfy eqs.~(\ref{eq.adiabatic_hot}) and (\ref{eq.adiabatic_cold}) in the quasistatic limit.

The cycle $2$ in fig.~\ref{fig_CA2} keeps $\tilde V_1$ and $\tilde V_3$ as the same volumes as the quasistatic cycles but $\tilde V_2$ and $\tilde V_4$ should be changed so as to satisfy the adiabatic curves $T_{hw}\tilde V_2^{\gamma-1}=T_{cw}V_3^{\gamma-1}$ and $T_{hw}V_1^{\gamma-1}=T_{cw}\tilde V_4^{\gamma-1}$ (see~\cite{RGRGAAB2018} for the similar construction).
Therefore, we have
\begin{eqnarray}
&\tilde V_2=\left(\frac{T_{cw}}{T_{hw}}\right)^{\frac{1}{\gamma-1}}V_3,\label{eq.tilde_V2_cyc2}\\
&\tilde V_4=\left(\frac{T_{hw}}{T_{cw}}\right)^{\frac{1}{\gamma-1}}V_1.\label{eq.tilde_V4_cyc2}
\end{eqnarray}
In this cycle $2$, because of $\tilde V_1=V_1$ and $\tilde V_3=V_3$, the compression ratio is held fixed as $\tilde V_3/\tilde V_1=V_3/V_1$.
Therefore, some parts of the CA cycle is out of the Carnot cycle as shown in $T$-$V$ or $T$-$S$ diagram in fig.~\ref{fig_CA2}. The work is given by
\begin{eqnarray}
W&&=Nk_{\rm B}(T_{hw}-T_{cw})\ln \frac{\tilde V_2}{V_1}\nonumber\\
&&=Nk_{\rm B}(T_{hw}-T_{cw})\left(\ln \frac{V_2}{V_1}+\frac{1}{\gamma-1}\ln \frac{T_hT_{cw}}{T_cT_{hw}}\right),\label{eq.W_2}
\end{eqnarray} 
where we used eqs.~(\ref{eq.adiabatic_cold}) and (\ref{eq.tilde_V2_cyc2}).
Note that eq.~(\ref{eq.W_2}) may be larger than the quasistatic work $W_{\rm qs}=Nk_{\rm B}(T_h-T_c)\ln (V_2/V_1)$ in some cases, depending on the values of $T_{hw}$ and $T_{cw}$.

The times taken for the isothermal processes of the cycle read
\begin{eqnarray}
&t_h=\frac{Nk_{\rm B}T_{hw}}{\alpha(T_h-T_{hw})}\ln \frac{\tilde V_2}{V_1}=\frac{Nk_{\rm B}T_{hw}}{\alpha(T_h-T_{hw})}\left(\ln \frac{V_2}{V_1}+\frac{1}{\gamma-1}\ln \frac{T_hT_{cw}}{T_cT_{hw}}\right),\label{eq.t_h_first}\\
&t_c=\frac{Nk_{\rm B}T_{cw}}{\beta(T_{cw}-T_c)}\ln \frac{V_3}{\tilde V_4}=\frac{Nk_{\rm B}T_{cw}}{\beta(T_{cw}-T_c)}\left(\ln \frac{V_2}{V_1}+\frac{1}{\gamma-1}\ln \frac{T_hT_{cw}}{T_cT_{hw}}\right),\label{eq.t_c_first}
\end{eqnarray}
where we used eqs.~(\ref{eq.adiabatic_cold}), (\ref{eq.tilde_V2_cyc2}), and (\ref{eq.tilde_V4_cyc2}).
These relations show that the time for each isothermal process depends on the steady temperatures of both sides of the isothermal processes as $t_h=t_h(T_{hw}, T_{cw})$ and $t_c=t_c(T_{hw}, T_{cw})$, which contrasts to eqs.~(\ref{eq.t_h}) and (\ref{eq.t_c}) for the cycle $1$ in fig~\ref{fig_CA1}.

Moreover, the work and the heat from the hot heat bath at the maximum power, $W^*$ and $Q_h^*$, also read
\begin{eqnarray}
W^*&=Nk_{\rm B}(T_{hw}^*-T_{cw}^*)\ln \frac{\tilde V_2}{V_1}\nonumber\\
&=\frac{(\alpha T_h)^{1/2}+(\beta T_c)^{1/2}}{\alpha^{1/2}+\beta^{1/2}}Nk_{\rm B}(T_h^{1/2}-T_c^{1/2})\left(\ln \frac{V_2}{V_1}+\frac{1}{\gamma-1}\ln \sqrt{\frac{T_{h}}{T_{c}}}\right),\label{eq.W_first}\\
Q_h^*&=T_{hw}^*\Delta S_h\nonumber\\
&=\frac{(\alpha T_h)^{1/2}+(\beta T_c)^{1/2}}{\alpha^{1/2}+\beta^{1/2}}T_h^{1/2}Nk_{\rm B}\ln \frac{\tilde V_2}{V_1}\nonumber\\
&=\frac{(\alpha T_h)^{1/2}+(\beta T_c)^{1/2}}{\alpha^{1/2}+\beta^{1/2}}T_h^{1/2}Nk_{\rm B}\left(\ln \frac{V_2}{V_1}+\frac{1}{\gamma-1}\ln \sqrt{\frac{T_{h}}{T_{c}}}\right),\label{eq.Qh_first}
\end{eqnarray}
respectively. It is obvious that the efficiency at maximum power of the cycle $2$ recovers the CA efficiency from eqs.~(\ref{eq.W_first}) and (\ref{eq.Qh_first}).

Meanwhile, the cycle $3$ in fig.~\ref{fig_CA3} keeps $\tilde V_2$ and $\tilde V_4$ as the same volumes as the quasistatic cycles but $\tilde V_1$ and $\tilde V_3$ should be changed so as to satisfy the adiabatic curves $T_{hw}V_2^{\gamma-1}=T_{cw}\tilde V_3^{\gamma-1}$ and $T_{hw}\tilde V_1^{\gamma-1}=T_{cw}V_4^{\gamma-1}$ (see~\cite{GKPR1978} for the similar construction).
Therefore, we have
\begin{eqnarray}
&\tilde V_3=\left(\frac{T_{hw}}{T_{cw}}\right)^{\frac{1}{\gamma-1}}V_2,\label{eq.tilde_V1_cyc3}\\
&\tilde V_1=\left(\frac{T_{cw}}{T_{hw}}\right)^{\frac{1}{\gamma-1}}V_4.\label{eq.tilde_V3_cyc3}
\end{eqnarray}
As shown in $T$-$V$ or $T$-$S$ diagram in fig.~\ref{fig_CA3}, by construction, the cycle $3$ is entirely inside the Carnot cycle.
The work of this cycle reads
\begin{eqnarray}
W&&=Nk_{\rm B}(T_{hw}-T_{cw})\ln \frac{V_2}{\tilde V_1}\nonumber\\
&&=Nk_{\rm B}(T_{hw}-T_{cw})\left(\ln \frac{V_2}{V_1}-\frac{1}{\gamma-1}\ln \frac{T_hT_{cw}}{T_cT_{hw}}\right),\label{eq.W_3}
\end{eqnarray} 
where we used eqs.~(\ref{eq.adiabatic_hot}) and (\ref{eq.tilde_V3_cyc3}).
The times taken for the isothermal processes of the cycle read
\begin{eqnarray}
&t_h=\frac{Nk_{\rm B}T_{hw}}{\alpha(T_h-T_{hw})}\ln \frac{V_2}{\tilde V_1}=\frac{Nk_{\rm B}T_{hw}}{\alpha(T_h-T_{hw})}\left(\ln \frac{V_2}{V_1}-\frac{1}{\gamma-1}\ln \frac{T_hT_{cw}}{T_cT_{hw}}\right),\label{eq.t_h_second}\\
&t_c=\frac{Nk_{\rm B}T_{cw}}{\beta(T_{cw}-T_c)}\ln \frac{\tilde V_3}{V_4}=\frac{Nk_{\rm B}T_{cw}}{\beta(T_{cw}-T_c)}\left(\ln \frac{V_2}{V_1}-\frac{1}{\gamma-1}\ln \frac{T_hT_{cw}}{T_cT_{hw}}\right),\label{eq.t_c_second}
\end{eqnarray}
where we used eqs.~(\ref{eq.adiabatic_hot}), (\ref{eq.tilde_V1_cyc3}), and (\ref{eq.tilde_V3_cyc3}).
These relations show that the time for each isothermal process depends on the steady temperatures of both sides of the isothermal processes $t_h=t_h(T_{hw}, T_{cw})$ and $t_c=t_c(T_{hw}, T_{cw})$, as is the same as the cycle $2$ in fig.~\ref{fig_CA2}.

The work and the heat from the hot heat bath at the maximum power, $W^*$ and $Q_h^*$, also read
\begin{eqnarray}
W^*&=Nk_{\rm B}(T_{hw}^*-T_{cw}^*)\ln \frac{V_2}{\tilde V_1}\nonumber\\
&=\frac{(\alpha T_h)^{1/2}+(\beta T_c)^{1/2}}{\alpha^{1/2}+\beta^{1/2}}Nk_{\rm B}(T_h^{1/2}-T_c^{1/2})\left(\ln \frac{V_2}{V_1}-\frac{1}{\gamma-1}\ln \sqrt{\frac{T_{h}}{T_{c}}}\right),\label{eq.W_second}\\
Q_h^*&=T_{hw}^*\Delta S_h\nonumber\\
&=\frac{(\alpha T_h)^{1/2}+(\beta T_c)^{1/2}}{\alpha^{1/2}+\beta^{1/2}}T_h^{1/2}Nk_{\rm B}\ln \frac{V_2}{\tilde V_1}\nonumber\\
&=\frac{(\alpha T_h)^{1/2}+(\beta T_c)^{1/2}}{\alpha^{1/2}+\beta^{1/2}}T_h^{1/2}Nk_{\rm B}\left(\ln \frac{V_2}{V_1}-\frac{1}{\gamma-1}\ln \sqrt{\frac{T_{h}}{T_{c}}}\right),\label{eq.Qh_second}
\end{eqnarray}
respectively. Also in this case, the CA efficiency as the efficiency at maximum power is recovered from eqs.~(\ref{eq.W_second}) and (\ref{eq.Qh_second}).

From the above calculations, we can understand why the cycle $1$ in fig.~\ref{fig_CA1}, among the three, may be considered the simplest cycle that realizes the CA cycle:
In the cycle $1$, the time taken for each isothermal process depends on each steady temperature as $t_h=t_h(T_{hw})$ and $t_c=t_c(T_{cw})$ (eqs.~(\ref{eq.t_h}) and (\ref{eq.t_c})).
In contrast, in the cycles $2$ and $3$ in figs.~\ref{fig_CA2} and~\ref{fig_CA3}, the time taken for each isothermal process depends on both steady temperatures as $t_h=t_h(T_{hw}, T_{cw})$ and $t_c=t_c(T_{hw}, T_{cw})$ (eqs.~(\ref{eq.t_h_first}) and (\ref{eq.t_c_first}) and eqs.~(\ref{eq.t_h_second}) and (\ref{eq.t_c_second})), which is more complicated than that in the cycle $1$.
This suggests that the cycle $1$ may be more easily realized than the other two from an experimental point of view.
Furthermore, the work and heat at the maximum power $W^*$ and $Q_h^*$ also take the simplest form for the cycle $1$ in fig.~\ref{fig_CA1}; compare $W^*$ and $Q_h^*$ of the cycle $1$ (eqs.~(\ref{eq.W_simplest}) and (\ref{eq.Qh_simplest})) with those of the cycle $2$ (eqs.~(\ref{eq.W_first}) and (\ref{eq.Qh_first})) and the cycle $3$ (eqs.~(\ref{eq.W_second}) and (\ref{eq.Qh_second})).

\subsection{Maximum power regime as Pareto front}\label{Pareto}
\begin{figure}[!t]
\begin{center}
\includegraphics[scale=0.7]{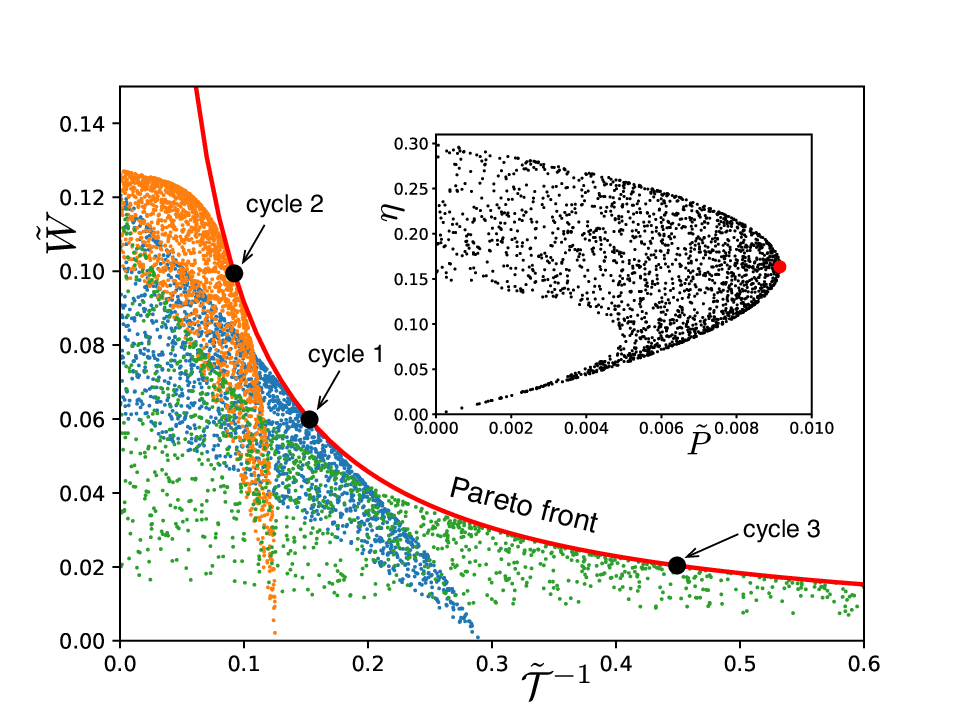}
\caption{The Pareto front $\tilde W=\tilde P^*/\tilde \mathcal T^{-1}$ (red solid curve) showing the best trade-off between the work and the inverse cycle-time (speed). Any point on the Pareto front corresponds to the maximum power regime of the CA cycle, where the three points (black-filled circles) correspond to the maximum power regimes of the cycles $1$--$3$ in figs.~\ref{fig_CA1}--\ref{fig_CA3}.
Meanwhile, the dots denote the various working regimes of the cycle $1$ (blue), the cycle $2$ (orange), and the cycle $3$ (green).
By sampling the nondimensionalized temperature differences $\tilde x$ and $\tilde y$ uniformly from $[0, (1-(T_c/T_h))/2]$, 
we plotted each dot $(\tilde \mathcal T^{-1}(\tilde x, \tilde y), \tilde W(\tilde x, \tilde y))$ of these three cycles using the same $\tilde x$ and $\tilde y$. Inset: the corresponding $\eta$-$\tilde P$ diagram, which is the same for the three cycles. The red-filled circle represents the maximum power regime $(\eta^*, \tilde P^*)$, which corresponds to the Pareto front in the main figure.
As the parameters, we used $\gamma=5/3$, $V_2/V_1=1.5$, $T_c/T_h=0.7$, and $\beta/\alpha=2$, which yields $\tilde W_{\rm qs}\simeq 0.121$.}\label{fig_Pareto}
\end{center}
\end{figure}

Although we have shown that the cycle $1$ may be the simplest construction of the CA cycle among the three in section~\ref{comparison}, here we show that they are equal to each other 
from a viewpoint of multi-objective optimization~\cite{D2001}.

Consider the problem of finding the best cycle constructions such that they increase the work $W$ and the inverse cycle-time $\mathcal T^{-1}$ ($\mathcal T\equiv t_h+t_c$) regarded as the speed of the cycle as multi-objective functions, where both quantities are practically favorable to be increased.
One possible solution to this problem is the so-called Pareto front that realizes the best trade-off between the objective functions on which increasing one without decreasing the other is impossible. Fortunately, from the results so far, we already know that in the maximum power regime the work and the inverse cycle-time of the cycles $1$--$3$ show the trade-off relationship in such a way that the maximum power $P^*$ is held fixed as eq.~(\ref{eq.Pmax}) (and $\eta^*$ as the CA efficiency).
So, we expect that the curve $W=P^*/\mathcal T^{-1}$ constitutes the Pareto front of this problem.

In fig.~\ref{fig_Pareto}, we have confirmed the above consideration by plotting the various working regimes of the cycles $1$-$3$ in figs.~\ref{fig_CA1}--\ref{fig_CA3} together with the Pareto front $\tilde W=\tilde P^*/\tilde \mathcal T^{-1}$, where we used the nondimensionalized work $\tilde W$, cycle-time $\tilde \mathcal T$, and maximum power $\tilde P^*$ defined as
\begin{eqnarray}
&&\tilde W\equiv \frac{W}{Nk_{\rm B}T_h},\\
&&\tilde \mathcal T\equiv \mathcal T/(Nk_{\rm B}/\alpha),\\
&&\tilde P^*\equiv P^*/(\alpha T_h),
\end{eqnarray}
respectively.
Then, defining the nondimensionalized temperature differences $\tilde x\equiv (T_h-T_{hw})/T_h$ and $\tilde y\equiv (T_{cw}-T_c)/T_h$,
$\tilde W$ and $\tilde \mathcal T$ of the cycles $1$-$3$ are expressed as the functions of $\tilde x$ and $\tilde y$ once the dimensionless quantities $\gamma$, $V_2/V_1$, $T_c/T_h$, and $\beta/\alpha$ are given.
The envelope of the working regimes correspondng to various $\tilde x$ and $\tilde y$ constitutes the Pareto front, and each of the three cycles at the maximum power corresponds to a different point on the Pareto front. 
Each point (``solution") on the Pareto front is equal to each other because it is a Pareto optimum such that increasing one, either work or inverse cycle-time in this case, is impossible without decreasing the other. The further choice depends on one's preference:
The cycle $2$ ``prefers" the work, while the cycle $3$ ``prefers" the time (speed), compared to the cycle $1$. 
This is also clear from $T$-$S$ diagrams in figs.~\ref{fig_CA1}--\ref{fig_CA3} as the work is given as the area enclosed by the cycle on $T$-$S$ diagram.
In principle, we may consider infinitely many constructions of the CA cycle whose maximum power regimes are on the Pareto front under the constraint that the CA cycle recovers the Carnot cycle with the predetermined $V_i$'s in the quasistatic limit.

The maximum power regime of the CA cycle is efficient not only because it outputs the maximum power but also it is Pareto optimal. Therefore, we may characterize the CA cycle in the maximum power regime as the preferred solutions realizing the best trade-off between the work and the inverse cycle-time.
This result is complementary to recent studies~\cite{GAGMRH2020,GARMH2020,MM2020,AS2021,ERAPLN2023} on the characterization of the performance of heat engines as a problem optimizing multi-objective functions including efficiency and power.

It should be noted that in the linear response regime $\Delta T \equiv T_h-T_c\to 0$, the works at the maximum power (eqs.~(\ref{eq.W_1}), (\ref{eq.W_2}), and (\ref{eq.W_3})) agree with the half of the quasistatic work $W_{\rm qs}=Nk_{\rm B}\Delta T \ln (V_2/V_1)$ in the first order of $\Delta T$:
\begin{eqnarray}
W^*\simeq \frac{1}{2}Nk_{\rm B}\Delta T\ln \frac{V_2}{V_1}=\frac{W_{\rm qs}}{2},\label{eq.half_Wqs}
\end{eqnarray}
which is consistent with the linear irreversible thermodynamics framework~\cite{VdB2005,IO2009}.
At the same time, the inverse cycle-time of the cycles $1$--$3$ is commonly given by
\begin{eqnarray}
\frac{1}{\mathcal {T^*}}\simeq \frac{\alpha \beta}{(\alpha^{1/2}+\beta^{1/2})^2}\frac{\Delta T}{2Nk_{\rm B}\bar T\ln \frac{V_2}{V_1}},\label{eq.half_time}
\end{eqnarray}
where $\bar T\equiv (T_h+T_c)/2$ is the average temperature of the heat baths.
Therefore, the different three points on the Pareto front collapsed to the one point $(W^*, 1/\mathcal {T^*})$ with eqs.~(\ref{eq.half_Wqs}) and (\ref{eq.half_time}) in the linear response regime. 
This behavior is crucially tied to the universality of the linear irreversible heat engines~\cite{VdB2005,IO2009}. Only in the nonlinear response regimes with respect to $\Delta T$, the one point splits into the three points as in fig~\ref{fig_Pareto}.

\subsection{General working substances}
Although we used the ideal gas as the working substance for the detailed constructions of the CA cycle as the simplest case, 
the present results can also be applied to general working substances; the original CA cycle does not depend on the specific choice of the working substance~\cite{CA1975}.

For general working substances, eqs.~(\ref{eq.V_dyna_h}) and (\ref{eq.V_dyna_c}) should be generalized as
\begin{eqnarray}
&&T_h-T_{hw}=\frac{1}{\alpha}\left(\frac{\partial S_h}{\partial V}\right)_{T=T_{hw}}\frac{dV}{dt} \quad (0 \le t \le t_h),\label{eq.V_dyna_h_gen}\\
&&T_{cw}-T_c=-\frac{1}{\beta}\left(\frac{\partial S_c}{\partial V}\right)_{T=T_{cw}}\frac{dV}{dt} \quad (0 \le t \le t_c).\label{eq.V_dyna_c_gen}
\end{eqnarray}
Then, by specifying the working substance and using its explicit form of the entropy, we can repeat the same calculations as we did in sections~\ref{Simplest construction of CA cycle} and~\ref{comparison}. For example, it is an easy exercise to perform the calculations for a van der Waals gas~\cite{LL2002}, which may be considered a model of real gas.

\section{Concluding perspective}
In summary, we gave detailed constructions of the CA cycle on the thermodynamic plane, by using the ideal gas as the working substance.
The simplest construction was analyzed and compared with the other two constructions in which the relationship between the steady temperatures of the working substance and the times for the isothermal processes take the more complicated forms than the simplest case. 
Meanwhile, these constructions were shown to be equal to each other as they are the elements of the Pareto front that achieves the best trade-off between the work and the inverse cycle-time. 
We hope that these results lead to the precise characterization of the CA cycle and contribute to the development of finite-time thermodynamics.

In the present work, we have neglected the times for the adiabatic processes and adopted the Newton's law of cooling with the constant heat-transfer conductances as the heat-transfer law so that they are consistent with the original setup of the paper by Curzon and Ahlborn~\cite{CA1975}. However, it is known that the changes of these setups considerably affect the efficiency at maximum power~\cite{GKPR1978,LL2002,IO2017,RGRGAAB2018,CY1989}, where the CA efficiency may no longer generally hold.
Therefore, it is of interest to compare cycle constructions under such general setups.

\section*{Acknowledgments}
This work was supported by JSPS KAKENHI Grant Numbers 19K03651 and 22K03450.

\end{document}